# Small Polaron Localization, Jahn-Teller Distortion and Defects in $La_{0.67}Ca_{0.33}MnO_3$: An Infrared Spectroscopy and X-ray Absorption Analysis.

NÉSTOR E. MASSA*

*Laboratorio Nacional de Investigación y Servicios en Espectroscopía Optica, Centro CEQUINOR-Departamento de Química and Departamento de Física, Universidad Nacional de La Plata, C.C. 962 , 1900 La Plata, Argentina,*

HÉLIO C. N. TOLENTINO

*Laboratório Nacional de Luz Síncrotron, CP 6192, 13083-970 Campinas, São Paulo, Brazil,*

HORACIO SALVA

*Comisión Nacional de Energía Atómica, Centro Atómico Bariloche and Instituto Balseiro, 8400 Bariloche, Rio Negro, Argentina,*

JOSÉ ANTONIO ALONSO, MARÍA JESÚS MARTÍNEZ-LOPE, MARÍA TERESA CASAIS

*Instituto de Ciencia de Materiales de Madrid, Consejo Superior de Investigaciones Científicas, Cantoblanco, E-28049 Madrid, Spain.*

**Running title: Small Polaron, Jahn-Teller distortion and defects in $La_{0.67}Ca_{0.33}MnO_3$**

* nem@dalton.quimica.unlp.edu.ar

Telephone and fax : 54 (0)221 424 0172

# ABSTRACT

We combine infrared reflectivity and EXAFS (Extended X-ray Absorption Fine Structure) techniques to study the solid solution $La_{0.67}Ca_{0.33}MnO_3$ prepared by different methods yielding samples with different insulator-metal transition temperatures ($T_{IM}$). While the small polaron analysis of the optical conductivity provides a natural description for the higher frequency reflectivity tail of conducting samples, our structural results are in accord with two non-equivalent sites in the insulating phase of good-quality samples. Those sites, one for the $Mn^{3+}$ Jahn-Teller distorted octahedra and another for the $Mn^{4+}$ ion, gradually turn into one dynamically averaged below the transition $T_{IM.}$ On the other hand, carriers screening antiresonances near infrared longitudinal optical modes, above $T_{IM}$, mirror thermal activated small polarons weakly smearing EXAFS oscillations. We associate this to the lack of $M^{3+}$,$M^{4+}$ explicit structure in the Mn *K*-edge absorption band. Extra octahedra, detected by EXAFS below $T_{IM}$ in higher resistivity samples, seem to be excluded of participating in the dynamics of the insulator-metal transition shifting $T_{IM}$ toward lower temperatures.

# INTRODUCTION

The subtle interplay of lattice and magnetic effects on the insulator-metal transition in manganese oxides, with a distorted perovskite or pyrochlore structure, is of current interest because it is intimately linked to the appearance of negative colossal mangetoresistance (CMR), the large change in resistivity at about the phase transition temperature $T_{IM}$ when an external magnetic field is applied. This temperature is optimized for $La_{0.67}Ca_{0.33}MnO_3$ at x~0.33, the largest number of $Mn^{3+}$ ions with one, and only one $Mn^{4+}$ near neighbor required for random composition in the lattice.[1]

$La_{0.67}Ca_{0.33}MnO_3$ is an orthorhombically distorted perovskite. In $t_{2g}$-$e_g$-Mn-ions the electrons in the $t_{2g}$ band are localized while the $e_g$ levels are hybridized with oxygen 2p states. Replacement of $La^{3+}$, e.g., by $Ca^{2+}$ holes in the $e_g$ band, thus creating a $Mn^{3+}$ ($t_{2g}^3e_g^1$)- $Mn^{4+}$ ($t_{2g}^3e_g^0$) mixed valence. Ferromagnetism is favored by the Hund's rule while the double exchange has been earlier proposed by Zener[2] as the basic mechanism for the phase transition. Carrier hopping depends on the relative alignment of the spin carrier to the localized $Mn^{4+}$ spin; when the two spins are aligned the carrier avoids the strong on-site Hund exchange energy and thus hops easily**.**

Millis et al proposed, in addition to Zener's mechanism, a lattice distortion factor involving localized charge carriers. This introduces an increase in the energy required for the charge to hop and it has been shown to be necessary to reproduce the measured $T_{IM}$'s.[3]

The main purpose of this paper, besides reporting on same sample transport, magnetic, EXAFS and infrared optical measurements of $La_{0.67}Ca_{0.33}MnO_3$ is investigating quantitative changes in the phonon assisted conductivity, above and below $T_{IM}$, as well as implications of "as prepared" samples on the driving double exchange mechanism. As pointed out in Hwang et al [4] discrepancies in the overall behavior of $La_{0.67}Ca_{0.33}MnO_3$ have been ascribed to chemical disorder, oxygen deficiencies, grain boundary effects, lattice constants, etc. Magnetic and structural properties depend critically on the preparation method and consequent oxidation state of the sample.[5] The possible role of phase separation in CMR materials has been recently reviewed by Dagotto et al.[6]

## SAMPLE PREPARATION AND CHARACTERIZATION

Our samples have been prepared under high pressure conditions and using the standard sol-gel method. In the first case, about 0.5 g of stoichoimetric mixture of oxides ($La_2O_3$, $CaCO_3$, $MnCO_3$) were thoroughly ground and put in a 8mm gold capsule sealed and placed in a cylindrical graphite heater. The reaction was carried out in a piston-cylinder press (Rockland Research Co.) at a pressure of 20 kbar at 1000 ºC for 20 minutes. The materials were then quenched to room temperature and the pressure was subsequently released.

For a second set of samples our solid solution $La_{0.67}Ca_{0.33}MnO_3$ was prepared from citrate precursors obtained by soft chemistry procedures, i.e., starting from stoichiometric amounts of analytical grade $La_2O_3$, $CaCO_3$ and $MnCO_3$ solved in citric acid. Then, the citrate solution was evaporated, dried at 120 ºC and slowly decomposed at temperatures up to 700 ºC in air. Finally, the precursor was annealed in air at 1000 ºC for 12 hrs. The powder was pressed into pellets : one of them was sintered at 1000 ºC for 12 hrs and the second one at 1200 ºC for 12 hrs.

The three samples were characterized by X-ray powder diffraction (XRD) for phase identification and to asses phase purity. XRD patterns were collected with Cu $K_a$ radiation in a Siemens D-501 goniometer controlled by a DACO-MP computer. The XRD patterns correspond, in all cases, to well cristallized monophased perovskites, showing the superstructure reflections corresponding to the well-known orthorhombic distortion, describable in the $D_{16}^{2h}$-Pbnm space group.

Characterization was also done by resistivity and magnetization measurements on the "as prepared" samples. Four point resistivity measurements were performed between 4 and 300 K (figure 1). The curves show sharp changes in the slope denoting insulator to metal-like behavior in the range between 100 K and 300 K. The resistivity of the sample made using the sol-gel techniques and annealed at 1000 ºC (figure 1,a) has a sharp insulating to metal like transition at about $T_{IM}$ ~180 K in a high resistive environment. Being the resistivity two orders of magnitude lower, the one prepared under pressure has $T_{IM}$ at approximately 254 K (figure 1, c). In addition, an intermediate behavior was found in the sol-gel sample sintered in air at 1200 ºC. Its resistivity is shown in figure 1,b. We relate the two maxima, at ~205 K and 257 K, to those peaking at $T_{IM}$ in figures 1a) and 1c), respectively, in an apparent two phase sample. It should be noted $T_{IM}$ at ~180 K and ~254 K seems to represent two temperature limits for which broad transitions in $La_{0.67}Ca_{0.33}MnO_3$ are often found. The overall resistivities are in agreement with published results.

Magnetization measurements of our samples, also shown in figure 1, were done with a Quantum Design SQUID magnetometer under 500 G. From this we remark : a) the sample with highest resistivity shows the onset for magnetic ordering only at lower temperatures than the sharp insulator-metal transition at ~180 K (figure 1a); the saturation magnetization (1.5 $\mu_B$/ Mn) is significantly lower than expected, which seems to be the manifestation of a partial spin-glass behavior, induced by the presumably high structural disorder of this sample,[ 7] b) in the intermediate case, figure 1b, although the resistivity suggests that there is a miscibility gap with two distinct ferromagnetic phases the susceptibility only shows ferromagnetic saturation at ~100 K; c) the sample with at $T_{IM}$ ~254 K has a ferromagnetic transition Tc at about that temperature (figure 1c) as expected for a good specimen.

Although our primary objective was on the local structure of the Mn ions we also performed a neutron powder diffraction (NPD) study for the sample annealed at 1000ºC, in order to evaluate the nature of the defects presumably present in the crystal structure, and responsible for the low $T_{IM}$ measured for this sample. A high resolution NPD pattern was collected at the D2B diffractometer of the ILL-Grenoble, with a wavelength of 1.594 Å. The NPD pattern was refined by the Rietveld method, considering the standard structural model for orthorhombically distorted perovskites (sp. gr. Pbnm). The atomic parameters after the refinement are included in Table 1. The refinement confirmed the presence of a single perovskite phase; a low discrepancy factor asses the quality of the fit ($R_{wp}$= 5.20%, $\chi^2$= 1.40). The refined occupancy factors for (La,Ca) and O show a significant deficiency at those positions, leading to the crystallographic formula $La_{0.65(1)}Ca_{0.32(1)}MnO_{2.90(3)}$. This corresponds to an average oxidation state of +3.24(1) for Mn. The defective nature of the crystal structure of this phase can explain the high resistivity and the lower $T_C$ observed: the presence of O vacancies distributed at random hinders the double exchange mechanism responsible for both transport and ferromagnetism; moreover the anionic and cationic vacancies can have a strong localizing effect on the charge carriers. Additionally, the oxidation state determined for this sample from electroneutrality criteria is significantly lower (0.24 holes per formula unit) than that necessary for the optimum electronic doping, of 0.33 holes per formula unit. A lower oxygen content implies higher resitivity, lower insulator-metal transition, lower Tc and lower saturation magnetization.[8,9,10,11,12]

# INSTRUMENTATION

Infrared reflectivity spectra in near normal incidence were taken between 30 and 10000 $cm^{-1}$ in a Bruker 113v FTIR spectrometer with samples mounted in the cold finger of a DN 1754 Oxford cryostat. We have used an Au mirror as reference. Pellets were made from samples with different $T_{IM}$'s, diluted in CsI or polyethylene, adequate for infrared transmission measurements. These data were useful to match the temperature dependent behavior of our reflectivities and for help in phonon frequency assignments.

EXAFS measurements with a high quality to noise ratio were recorded using the XAS beam line installed at the 1.37 GeV storage ring of the Laboratorio Nacional de Luz Sincroton at Campinas, Brazil.[13] The synchrotron radiation was monochromatized by a channel-cut Si(111) crystal monochromator with a resolution of about 2 eV at the Mn K-edge (6539 eV). The resolution was about 1.5 eV for measurements around the Ca K-edge (4038 eV). The spectra were recorded in transmission mode, up to the momentum transfer of $2k=30Å^{-1}$, using two air-filled ion chambers as detectors. For this purpose, portions of the samples used in our infrared measurements were reduced to fine powder in an agate mortar. The fine-ground powder washed with alcohol was then uniformly spread onto nitro-cellulose filter, resulting, after evaporation, in a uniformly thick blackish film. For these measurements our three samples, corresponding to the resistivity behavior shown in figure 1, were mounted simultaneously in the cold finger of a CRYOMECH closed cycle He refrigerator, allowing measurements between 10 and 300K.

## REFLECTIVITY MEASUREMENTS-ANTIRESONANCES IN OXIDES

The dominant features in the infrared reflectivity at room temperature are main vibrational bands. They are zone center phonons given in the ideal cubic perovskite by

$$\Gamma(O_h^1) = 4F_{1u}+F_{2u} \qquad (1)$$

i.e., a triple degenerate acoustical mode and three infrared triple degenerate. The overall irreducible representation of our solid solution, a deviation of the already distorted cubic perovskite, is usually accepted as orthorhombic, $D_{2h}^{16}$ -Pbnm space group, with four formula units in the unit cell.[14] Adding Raman and infrared active modes of $La_{0,67}Ca_{0.33}MnO_3$, the predicted zone center phonons are

$$\Gamma(D_{2h}^{16})=8A_u+9B_{1u}+7B_{2u}+9B_{3u}+7A_g+5B_{1g}+7B_{2g}+5B_{3g}. \qquad (2)$$

It is always useful, however, to keep in mind that with x=0.33 our system does not have translational symmetry due to the intrinsic mixed crystal disorder.[15] This and the lack of an inversion center relax the gerade-ungerade selection rule effectively reducing the phonon coherent length, and thus, allowing partial optical detection of the vibrational density of states.[16]

We analyzed our reflectivity spectra simulating infrared-active features with damped Lorentzian oscillators in a classical formulation of the dielectric function. We added one plasma contribution (Drude term) to the dielectric simulation when the number of carriers required it.[17]

The temperature dependent reflectivity of the sample with $T_{IM} \approx 180$ K is shown in figure 2a. In these spectra, there are three bands associated to vibrational groups centered at ~175 $cm^{-1}$, ~350 $cm^{-1}$ and ~580 $cm^{-1}$ and a broad band at 800-1000 $cm^{-1}$ that we will see assigned to the small polaron binding energy.

Those three phonon frequencies correspond to envelops of the above predicted perovskite vibrational groups allowed in the space group $D_{2h}^{16}$-Pbnm. They are well defined at room temperature and may be easily followed into the metallic phase hinting that in this sample the fewer number of carriers does not achieve complete phonon screening even at 77 K. The overall phonon spectra agree with those reported by Kim et al.[18]

Interestingly, those non-totally screened phonons bring up the disclosure of incipient antiresonances at ~218 $cm^{-1}$, ~461 $cm^{-1}$, ~696 $cm^{-1}$ (figure 2 b) at temperatures above $T_{IM}$. They result from $e_g$ electrons interacting with longitudinal optical modes (and their macroscopic field within a highly polarizable lattice created by oxygen ions).[19] This localized effect, recreating an environment of strong electron-phonon interactions, is almost diluted at 160 K, below $T_{IM} \sim 180$ K at the onset temperature of the Drude tail. Although it is found in a small temperature range that behavior is characteristic of an ionic insulator i.e., phenomenological damping of transverse optical modes is reduced as the temperature lowers while close to the longitudinal modes there is a decrement in the reflectivity. We have already reported this behavior in several other semiconducting oxides such as $RNiO_3$ (R = Rare Earths)[20] as well as in pyrochlore $Tl_2Mn_2O_7$ [21] ,in this last one , with Raman degenerated infrared longitudinal modes.[22]

In metallic $LaNiO_3$ [17] extra absorptions on the Drude band near longitudinal optical frequencies reminisce those found of high Tc compounds. As recently pointed by Homes et al[23] these effects might be also related to charge fluctuations seen in lower dimensional conductors. The photoinduced infrared spectra of $K_{0.3}MnO_3$, a quasi-one dimensional conductor that undergoes a Peierls transition at 180 K, [24] reveals the strong nature of the electron-phonon interactions, i.e., the appearance of strong antiresonance absorptions at near optical modes frequencies in consonance with phonon interference with an electronic continuum. Within this view, in oxides, phonons couple with electrons localizing charges in other fashion than the expected for classical metals, these last ones with shorter screening lengths, with lattice regions with higher carrier density than in those in which they are localized.[23] We will see below that our small polaron analysis individualizes phonon groups and degree of localization that play a determinant role in the oxides conductivity.

Decreasing temperature, in $La_{0.67}Ca_{0.33}MnO_3$ ($T_{IM}$~180 K), the number of effective carriers ($N_{eff}$) [17] are only sufficient to smooth the sharp antiresonances of the insulating phase not developing, at 77 K, a well defined Drude edge and complete phonon screening. The infrared tail going down in intensity as the frequency increases and extending up to 10000 $cm^{-1}$ is a common denominator for all conducting oxides also present in our spectra.[25] Our fitting results are shown for $La_{0.67}Ca_{0.33}MnO_3$ ($T_{IM}$~180 K) in table 2. Using $m^*=10\, m_0$, where $m_0$ and $m^*$ are the free and effective electron mass, a rough estimation of $N_{eff}$ at 77 K yields~$10^{17}$, that is, the expected value for a semiconductor.

The infrared spectra of the intermediate sample with resistivity peaking at 205 K and 250 K has a defined weak Drude edge at about 100 $cm^{-1}$ showing, in agreement with its higher dc conductivity (figure 1b), an increment in the carriers mobility. Ferromagnetism saturation brings up the influence of spin ordering in the conductivity. On the other hand, this sample has a mixed phase character and its infrared spectrum does not have other peculiar feature than the already mentioned Drude edge.

The infrared reflectivity of $La_{0.67}Ca_{0.33}MnO_3$ prepared under high pressure ($T_{IM}$~254 K) **(**figure 3) at ambient temperature is dominated by phonons and a relative broad band associated with small polaron localization at higher frequencies. In addition, an incipient continuum at room temperature, on which all features appear, is consonant with a two order reduction in the bulk resistivity and ferromagnetism saturation at $T_{IM}$~ Tc~254 K. This, on the other hand, blends into the continuum the observation of sharp longitudinal optical antiresonances. When cooling, and although main phonon bands are still delineated, the carrier constituent grows in intensity and at difference of the other cases, fully develops a defined plasma edge yielding at 77 K $N_{eff}$~$10^{18}$.(figure 3). On the other hand, internal modes peak position clearly shifts below $T_{IM}$ to higher frequencies, i.e., involving Mn-O motions these vibrations are the likely affected by the polaron delocalization [26]. Fitting parameters for this case are shown in table 3.

## OPTICAL CONDUCTIVITY AND DISCUSSION

We interpret the experimental optical conductivity in terms of Reik and Heese theory for small polarons in the finite temperature approximation.[27] Thus, the real part of the optical conductivity for finite temperature, $\sigma_1(\omega,\beta)$ is given by

$$\sigma_1(\omega,\beta) = \sigma_{DC} \frac{sinh\left(\frac{1}{2}\hbar\omega\beta\right)\exp\left[-\omega^2\tau^2 r(\omega)\right]}{\frac{1}{2}\hbar\omega\beta\left[1+(\omega\tau\Delta)^2\right]^{1/4}}, \qquad (4)$$

$$r(\omega) = \left(\frac{2}{\omega\tau\Delta}\right)\ln\left\{\omega\tau\Delta + \left[1+(\omega\tau\Delta)^2\right]^{1/2}\right\} - \left[\frac{2}{(\omega\tau\Delta)^2}\right]\left\{\left[1+(\omega\tau\Delta)^2\right]^{1/2} - 1\right\}, \qquad (5)$$

with $\Delta = 2\varpi\tau$ (6)

and $\tau^2 = \frac{\left[sinh\left(\frac{1}{2}\hbar\varpi\beta\right)\right]}{2\varpi^2\eta}$. (7)

$\sigma_1(\omega,\beta)$, $\beta=1/kT$, is mainly three parameter dependent; $\sigma_{DC} = \sigma(0,\beta)$, the electrical conductivity (taken from our resistivity measurements); the frequency $\varpi_j$ that corresponds to the average between the transverse and the longitudinal optical mode of the $j^{th}$ restrahlen band; and $\eta$, that it is a parameter characterizing the strength of the electron-phonon interaction, i.e., the average number of phonons that contribute to the polarization around a localized polaron. From them, the only actual parameter free to fit for each phonon contribution is $\eta$ because phonon frequencies are fixed by reflectivity and transmission measurements.[28] We allow to vary $\varpi_j$, the average reststrahlen frequency, and $\eta_j$, the strength of the electron-phonon interaction for a particular vibrational group against the experimental data.

In manganites, one achieves agreement with the data if independent contributions of lattice and stretching vibrations are considered. The higher frequency tail originates from overtones or combination phonon sum processes. For each vibrational group the total number of phonons, $\xi$, involved in the conductivity is the sum of the number of phonons emitted and absorbed. At frequencies higher than $\eta\omega_0$, ($\omega_0$ an arbitrary phonon, $\eta$ average number of optical phonons of the polarization around a localized polaron) energy conserving processes are only possible for $\xi > \eta$ [29] restraining optical absorptions to a decrease as we increase the probing frequency, i.e., light cannot do better than totally strip the polarization of the polaron in the hopping process. Consequently, a band strength diminishes asymptotically as the frequency increases beyond lattice features giving origin to the infrared reflectivity tail. In semiconductors, like $LaNi_{0.50}Fe_{0.50}O_3$ [17] the simple hypothesis of assuming three independent contributions, one for each main perovskite vibrational group (average of reststrahlen bands), is enough to reproduce the spectrum range up to 10000 $cm^{-1}$. When there is an increment in the number of carriers in $LaNi_{1-x}Fe_xO_3$ the phonon contributions tend to blend in a kind of vibrational average of allowed phonons. The rest correspond to overtone and third order sum processes. We experimentally verified the number of phonons that form the polaron cloud increases as the phonon order**.** Table 4 shows that in $La_{0.67}Ca_{0.33}MnO_3$ ($T_{IM}$~180 K) the temperature dependent insulating behavior in the sample is mirrored in the evolution of $\eta$. The highest $\eta$ is associated with lattice phonons at ~170 $cm^{-1}$. It diminishes as the mobility increases suggesting a crossover to a more delocalized carrier regime but still keeping the small polaron character. This is supported by breathing phonons peaking at ~580 $cm^{-1}$ that are associated in the whole temperature range to electron-phonon strengths typical for a perovskite environment[29]

The analysis for $La_{0.67}Ca_{0.33}MnO_3$ (Tc~$T_{IM}$~254 K) follows the results of our first sample. Phonon bands at room temperature are not only better defined but splittings seen at ~300 $cm^{-1}$ suggest that this well defined substructure is due to normal ($Mn^{4+}$) and Jahn-Teller distorted ($Mn^{3+}$) octahedras, resolved beyond the intrinsic disorder of our perovskite with symmetry lower than cubic. The optical conductivity for breathing modes is in this case better adjusted considering two close-but-non-equivalent gaussians that are interpreted as due to sublattices with slight different environments with the same the phonon strength $\eta_j$**.** Figure 3 shows a Drude edge fully developed at about 100 K, the $\eta_j$'s reduced, in agreement with transport measurements, reflecting the lose in localization. The small polarons grow "larger", i.e., gradually delocalize, pass $T_{IM}$ and down to temperatures in which there is a more isotropic environment with spins accompanying the lattice polaron distortion.[30] As we cool down (Figure 3,Table 3) the 300 $cm^{-1}$ band merge into one broader feature due to a dynamic screening by the carriers. We associate this temperature to the last inflection of the resistivity at ~90 K in figure 1c. Significantly, this is the temperature brought up recently by scanning tunneling spectroscopy [31] at which a finite density of states at the Fermi level was recorded [32] In this sample octahedral internal modes are the only relevant phonons helping electron hopping in agreement with the association between distorted octahedra and polaron formation. And as we move toward higher frequencies the small polaron analysis again shows that there is an increment in the number of phonons contributing phonon summation processes constituting the polaron cloud (Figure 5, Table 5).

A further comment on this analysis is that the DC conductivity, $\sigma_{DC}$,, enters as a factor. Regularly, the value used here is from the same sample resistivity. We have found that in lower conductivity samples this contribution is much smaller than the one extrapolated from our optical conductivity prompting us to conclude that, in agreement with earlier findings in thin films, defects seem to play a significant role in downgrading the DC conductivity while the infrared response is related more realistically to the carrier dynamics.[33] On the other hand, in the sample with $T_{IM}$~254 K this difference is not significant since it is well within the uncertainty of an infrared measurement from a polycrystalline specimen. Note that resistivities of the same sample at 300 K and 80 K, figure 1, are about the same order while the infrared reflectivity shows remarkable changes as carriers become mobile.

## X-RAY ABSORPTION SPECTROSCOPY

From the above, it is transparent that the relation carrier-lattice-vibration is of primordial importance in the dynamics and conductivity of oxides and that local structure information plays an important role in elucidating the possible scenarios. Accordingly, in this section we exploit the fact the EXAFS (extended x-ray absorption fine structure) is a technique that allows independent knowledge bound to a particular ion. This is pertinent for the constituents of a mixed crystal [15] since the more subtle information on distances and nearest neighbors is otherwise missed in diffraction measurements. As pointed out by Subias et al[34] those differences are not observed by X-ray diffraction because the mixed nature of the lattice ions imposes resolution limits on lattice line half-widths. In fact, it is worth mention again that our X-ray diffraction patterns at 300 K do not yield any difference for the three cases. While this last technique gives the average structure, EXAFS is a tool that does not require long range crystal symmetries.[35]

We studied our samples primarily putting emphasis on the insulating-metal oxide transition and its relation to oxygen coordination shells. Counting on the nominal composition for the three cases in $La_{0.67}Ca_{0.33}MnO_3$, we now searched on structural differences against the variation on $T_{IM}$ and relate to resistivities, susceptibilites and optical measurements (figs. 1, 2, 3, 4) assimilating the small polaron analysis to Jahn-Teller type octahedral distortions.

The Mn K-edge EXAFS spectra for the three samples were collected at temperatures of 10K, 90K, 160K, 250K and 300K. At the Ca K-edge, only the sample with the highest $T_{MI}$ was recorded at 10K and 300K. The Mn K-edge spectra of one of the samples ($T_{MI}$≈254K) for three different temperatures are shown in figure 6. The spectral shapes are very similar among all samples and only subtle differences can be distinguished. These differences are better identified through the Fourier Transform analysis. For doing so, the EXAFS spectra were reduced by the standard procedure [35], using the WinXAS code.[37]. The oscillatory part of the EXAFS spectra were extracted by fitting a polynomial function to the monotonous decay, converted from energy into k-space, through the relation $\hbar k=(2m(E-E_K))^{1/2}$, where $E_K$ is the edge energy, and then the EXAFS signals, $\chi(k)$, weighted by $k^2$ to emphasize the large k contribution, were Fourier transformed from 2.7 up to 15 $Å^{-1}$. The Fourier Transform gives radial distribution function (RDF) modified mainly by the phase shifts, due to the absorber and backscatter potentials, and by backscattering amplitudes. It is normally called pseudo-RDF and contains the contribution of each nearest shell of atoms around the absorber. Figure 7(a,b) displays the $k^2$-weighted Fourier Transform for the sample with the highest $T_{MI}$ for three temperatures. The contribution from the coordination shell of oxygen atoms can be easily separated, within 0.9 Å to 2.3 Å, and then backtransformed into k-space. More distant shells are difficult to separate and their analysis involves taking into account complex multiple scattering paths [38,39]

Once the coordination shell is separated, structural parameters like bondlengths, coordination number and relative disorder, (R, N, $\Delta\sigma^2$) are obtained from least square fittings in k-space, using either experimental or theoretical phase shifts and back scattering amplitudes functions. We choose to analyze the data using theoretical phase and amplitudes functions extracted from FEFF6 code.[40] All the results presented here are corrected by the amplitude reduction factor $S_o^2$, an overall attenuation that is not accounted for in FEFF6, whose value is set equal to 0.7. We could have fixed the total coordination number equal to 6 and let $S_o^2$ to vary. However, the important point here is the ratio of the coordination number for each distance and how they change with temperature.[36,39,41]

It is worth mentioning that, attempting to have an experimental standard for a better first shell octahedral analyses, we also prepared with a solid state reaction under high pressure, the two end compounds ($LaMnO_3$, spontaneous $Mn^{3+}$ Jahn-Teller elongation, and $CaMnO_3$, $Mn^{4+}$). We realized that both materials proved to be poor standards due to the octahedral intrinsic deformation in the distorted perovskite lattice and we thus discarded both. In what follows, in agreement with earlier conclusions by Li et al,[35] we discuss our results using only the theoretical standard for the oxygen coordination shell fitting.

# EXAFS RESULTS AND DISCUSSION

Figure 7a and 7b shows the modulus and imaginary parts, respectively, of the Fourier transform of the $k^2$-weighted Mn K-edge EXAFS oscillations for $La_{0.67}Ca_{0.33}MnO_3$ ($T_{IM} \approx 254$ K). Main temperature dependent features associated with Mn-O, Mn-La/Ca, Mn-O-Mn distances are in very good agreement with results by Lanzara et al for $La_{0.75}Ca_{0.25}MnO_3$ [42]. They corroborate earlier results hinting that in Ca-doped $LaMnO_3$ temperature dependent nearest neighbors Mn-O differences were consistent with small polaron Jahn-Teller distortions.[43] Overall attenuation is observed as the temperature increases, owing to the increase of the thermal disorder.

The EXAFS amplitude of the Mn-O contribution steadily increases from 300 to 10K. The feature at ~1.7Å, seen in the imaginary part, almost doubles from room temperature to 10K. Here, it should be pointed out that while the overall features decrease in amplitude when temperature raises, there is an increase in amplitude at ~1.9Å. Once corrected by the phase shift, this feature can directly associated to the appearance of the second Mn site. The Mn-La/Ca and Mn-O-Mn contributions seems to present a transition between a room temperature and very low temperature (<100K) behavior. It is known that these features are sensitive to the photoelectron multiple scattering paths and, as consequence, to the octahedral arrangement and distortion. Qualitatively speaking, the low temperature structure points to a less distorted situation.

Our quantitatively analysis, through the Fourier filtering technique previously described, focused on the oxygen coordination shell bounds. The analysis approach was to simulate the coordination shell using, as a first step, two different distances, constraining the basal distance close to the value of 1.92Å and letting the second distance to vary freely. The second step was to include a third distance and allow further refinement of the octahedral elongation distances. The basal octahedral and elongation distances are in excellent agreement with earlier known results.[42] Table 6 shows our results approaching the octahedral fit with two and three shell hypotheses. While at 300, 250 and 160 K it is necessary to use a three shell surround, the measurements at 90 K and 10 K are satisfied with only two. We, nevertheless, quote our three shell fit for these two last temperatures emphasizing the fact that no new information is obtained beyond two shell distances, i. e., 1.92 Å for the base and 2.04 Å (2.05 Å) for the apex ion.

It is also illustrative recalling that our samples, even at room temperature, have a thermally excited polaronic semiconducting behavior. We commented above that this correlates to longitudinal modes interacting with semilocalized carriers weakening the field splitting longitudinal and transverse modes. Thermally activated small polarons may thus be visualized as a rarified cloud in the inner octahedral region where the EXAFS electron-wave generates oscillations now weakly smeared. This, in turn, results in a smoothed break in the $k^2$ weighted Mn-K edge profile associated to a possible feature expected as consequence of the "mixture" of $M^{3+}$, $M^{4+}$ ions.[44] Polarons are expected to be static on a time scale of 1ps but they are surely hopping.[45]. Our findings support the image on charge spread over several ions with a net increase in the mixed system covalency as have been already put forward by Lu et al [46] and Mizokawa and Fujimori [47] as well as the possible hybridization of Mn 4p and 3d and O2p states.[43] For these systems bond admixing was also earlier proposed by Croft el al.[48]

We then conclude that for $La_{0.67}Ca_{0.33}MnO_3$ ($T_{IM}$≈254 K) at room temperature we have two distinct sub-lattices, one of them associated to the local Jahn-Teller distortion (d~2.15 Å). The elongation, consequence of the outer $Mn^{3+}$ 3d level split, compatible with Mn ion charge disproportion, being the source of the small polaron scenario for an intrinsic static structural octahedral disorder in manganites. This situation persists down to $T_{IM}$, below which, entering in the interval of the colossal magnetoresistance behavior (~100K), the picture of two non-equivalent octahedra turns gradually into one prevailing site where now the electron move fast enough.[8]

A more quantitative picture of octahedral polaronic deformation is in figure 8, where it is shown the separation of bonds distances in the radial distribution function for Mn-O extracted from EXAFS oscillations. It provides the amplitude of the local distortions associated to them. As we increase the temperature a longer distance at about 2.15Å appears as a more localized distortion in which the double exchange interaction is not effective.

Although presenting more peaks corresponding to many different near neighbors, the Ca K-edge Fourier transform $k^2$ weighted EXAFS oscillations for this sample, shown in figure 9, confirms the overall continuous character of the $T_{IM}$ phase transition. It does not present major structural changes in going from 10 to 300 K. The overall attenuation is just coming from the increase of thermal disorder contribution.

The fit to experimental first shell filtered oscillations for $La_{0.67}Ca_{0.33}MnO_3$ in the intermediate case, with $T_{IM1}$≈205 K and $T_{IM2}$≈257 K, allows to visualize dynamical screening effects due to the increment in the number of carriers (Table 7). This is found in the lower temperature resistivity decrease (figure 1b) and the consequent appearance of a weak, but well-defined, low frequency Drude edge. The breakdown yields two non equivalent sites down to 160 K, and one low temperature average site at 10 K, indicating that the dynamical process imposed by the double exchange interaction prevents low temperature discrimination of the intervening sublattices and acquainting in the conductivity the decrease of spin disorder close to the onset of the ferromagnetic saturation at ~100 K. (fig. 1b). In addition we can not rule out that, between 180 K and 254 K, our sample might have charge dynamical fluctuations as result of oxygen rearrangements and microstrains originating in the competition between the retreating paramagnetic microdomains. This gradual slight changes in the octahedra reflects changing from an static picture to a dynamic one, and the increasing carrier path by the size increment of the ferromagnetic clusters.[49]

Strain dependence has been singled out as a main factor on determining $T_{IM}$ in manganites. Biaxial stress further provided evidence in favor a relating the CMR to Jahn-Teller electron-phonon coupling.[50] Large in-plain strain seem to exhibit large local Jahn-Teller distortions. Strains induces lattice defects trapping the $e_g$ electrons in local distortions and, thus, enhancing the resistivity.[51]

Finally, in our sol-gel sample $La_{0.67}Ca_{0.33}MnO_3$ ($T_{IM}$≈180 K), we have the same overall temperature dependent behavior as in the sample prepared under high pressure but with the main difference that a sublattice distortion remain identifiable down to 10K (Table 8). The lattice may have regions in which the double exchange interaction is inoperative, and, as in the case of $LaMnO_{3+0.15(0.07)}$ [52], we interpret the absence of a metal oxide behavior in portions of the lattice as distinctive octahedra sets related to intergrain walls. Stronger small polaron localization would be found adjacent to other areas in which a delocalized polaron scenario would be a more appropriate description and, thus, having regions conducting more than orders. An excess of oxygen resulting from La and Ca vacancies corresponds to $Mn^{4+}$ percentages not being optimized in their link to $Mn^{3+}$ 5[53]. On the other hand, the defect chemistry in parent compound $LaMnO_{3+\delta}$ is better described by randomly distributed La and Mn vacancies in equal amounts rather than that with oxygen excess intertitials. Thus, a relatively high proportion of Mn vacancies perturbs the connecting paths for the transport of holes across Mn-O-Mn. We may then think that it is the presence of random vacancies having strong localizing effects on the available charges, preventing them to move across the crystal, what makes a higher resistivity sample.[54] At the same time, the defective character of this sample can hinder the perfect ordering of spins and thus reduce the saturation magnetization, inducing, at low temperatures, a spin-glass like behavior affecting to a significant fraction of the total magnetic moments.

# CONCLUSIONS

Summarizing, we have studied samples of $La_{0.67}Ca_{0.33}MnO_3$ that have an explicit single and multiphase behavior in their conductivity using infrared and EXAFS techniques. We have shown that these measurements yield information that strongly support the Jahn-Teller distortion as small lattice polaron that delocalizes below $T_{IM}$. Further, we remark that our small polaron binding energies calculated by $E_b \leq \varpi\eta/2$ [27] and assigned to a broad infrared band in the range from 800 to 2500 $cm^{-1}$, are in agreement with the values deduced from resistivity measurements within the framework of the Holstein model for small polaron in molecular materials.[54]

The resistivitity decrease at low temperatures is linked to the gradual appearance of an average octahedral site denoting an effective double exchange mechanism in an strong electron-phonon environment in which, prevailing breathing phonons, all vibrational modes play individual significant roles in the conductivity. The small polaron analysis of the infrared optical conductivity suggests oxides are materials that do not bulk conduct uniformly.

EXAFS provides structural information on the small polaron localizing temperatures higher than $T_{IM}$. Our results are in agreement with two non equivalent sites in good samples. Both, one for the $Mn^{3+}$ Jahn-Teller distorted octahedra and another for the $Mn^{4+}$ ion gradually turn into one dynamically averaged site below the transition $T_{IM}$. We find octahedral distances that are in full agreement with those already reported in the literature.[42]

Exceptions of that behavior seem to be created when there are stoichiometric defects in portions of the sample. This situation, in turn, is identifiable by distinctive sublattices at temperatures lower than $T_{IM}$ down to 10 K. An identification of the possible existence of lattice mismatching is suggested by octahedra that seem to be excluded from participating in the dynamics of the insulator-metal transition due to chemical and/or structural reasons. This shifts the insulating to metal like transition toward lower temperatures and brings up coexisting insulating and more conducting lattice environments yielding ill resolved transport pictures for which our intermediate sample (figure 1b) is an extreme case. We interpret this effect, defining a criterion on sample quality, as the origin of an overlapping broad background found in some resistivity measurements at temperatures lower than $\sim Tc \sim T_{IM}$.

This research has been partially performed at the LNLS (Laboratório Nacional de Luz Síncrotron) (project XAS 475/99) to whom N. E. M. also thanks for the hospitality. Partial support was also obtained from grant PICT Nº 8017 of the Argentinean CONICET (Consejo Nacional de Investigaciones Científicas y Técnicas). J. A. A., M. J. M-L and M. T. C. acknowledge the financial assistance of the Ministerio de Ciencia y Tecnología under Project Nº MAT2001-0539.

## FIGURE CAPTIONS

**Figure 1.** Resistivity and susceptibility measurements of $La_{0.67}Ca_{0.33}$ $MnO_3$. a) sample prepared using sol-gel techniques; b) same as in a) but temperature cycle once at 1200 ºC; c) sample prepared at 20kbar and 1000 ºC. Dots signal the temperatures at which the reflectivity was measured.

**Figure 2. (a)**Temperature dependent reflectivity spectra of $La_{0.67}Ca_{0.33}$ $MnO_3$ prepared with sol-gel techniques ($T_{IM}$ ~180 K).Dots : experimental; full lines : fitted.

**Figure 2 (b)** Incipient antiresonances in the reflectivity spectra of $La_{0.67}Ca_{0.33}$ $MnO_3$ prepared with sol-gel techniques ($T_{IM}$ ~180 K) at near longitudinal optical mode frequencies.

**Figure 3.** Temperature dependent reflectivity spectra of $La_{0.67}Ca_{0.33}$ $MnO_3$ prepared in a solid state reaction under 20kbar (Tc~$T_{IM}$~254K).Dots : experimental; full lines: fitted.

**Figure 4**. Temperature dependent real part of the optical conductivity of $La_{0.67}Ca_{0.33}MnO_3$ ($T_{IM}$ ~180 K). Diamond upper traces are optical conductivities drew from their respective reflectivities. The superposing full lines are estimates using equation 4**.** Individual gaussians correspond to different phonon contributions discussed in the text.

**Figure 5**. Temperature dependent real part of the optical conductivity of $La_{0.67}Ca_{0.33}MnO_3$ ($T_c$~$T_{IM}$~254 K). Diamond upper traces are optical conductivities drew from their respective reflectivities. The superposing full lines are estimates using equation 4. Individual gaussians correspond to different phonon contributions discussed in the text.

**Figure 6**. Mn K-edge EXAFS spectra of $La_{0.67}Ca_{0.33}$ $MnO_3$ ($T_c$~$T_{IM}$ ~254 K) for three different temperatures.(full line, 10K; dashes, 160K; and dots, 300K).

**Figure 7 a.** Temperature dependence of the FT($k^2\chi$(k)) modulus for the Mn K-edge of $La_{0.67}Ca_{0.33}$ $MnO_3$ ($T_c$~$T_{IM}$ ~254 K).

**Figure 7 b.** Temperature dependence of the FT($k^2\chi$(k)) imaginary part .for the Mn K-edge of $La_{0.67}Ca_{0.33}$ $MnO_3$ ($T_c$~$T_{IM}$ ~254 K).

**Figure 8**. Mn-O pair distribution function of $La_{0.67}Ca_{0.33}$ $MnO_3$ ($T_c$ ~ $T_{IM}$ ~254 K) form the EXAFS oscillation analysis of Mn-O bonds lengths..

**Figure 9** Ca K-edge FT($k^2\chi$(k)) for $La_{0.67}Ca_{0.33}MnO_3$ ($T_c$~$T_{IM}$~254 K). Note that the overall structure does not change in going from 300 to 10K.

# Table 1

Positional and thermal parameters at 300 K for $La_{0.67}Ca_{0.33}MnO_3$, annealed at 1000ºC (refined in the orthorhombic Pbnm space group, Z= 4, from NPD data.) as well as main interatomic distances (Å) and angles (º) concerning the $MnO_6$ octahedra

| Atom | site | x | y | z | $f_{occ}$ | B(Å$^2$) |
|---|---|---|---|---|---|---|
| La | 4c | 0.996(1) | 0.0168(8) | 0.25 | 0.65(1) | 0.62(6) |
| Ca | 4c | 0.996(1) | 0.0168(8) | 0.25 | 0.32(1) | 0.62(6) |
| Mn | 4b | 0.5 | 0 | 0 | 1.0 | 0.57(9) |
| O1 | 4c | 0.1061(5) | 0.4516(5) | 0.25 | 0.91(3) | 0.30(13) |
| O2 | 8d | 0.6985(3) | 0.2988(3) | 0.0563(2) | 1.01(3) | 1.16(8) |

| | | | | | |
|---|---|---|---|---|---|
| Mn-O1 | 1.9591(8) Å | x2 | Mn-O1-Mn | 159.83(4)º | |
| Mn-O2 | 1.949(4) Å | x2 | Mn-O2-Mn | 161.21(16)º | x2 |
| | 1.970(4) Å | x2 | | | |

Reliability factors: $R_p$= 4.14%, $R_{wp}$=5.20%, $R_{exp}$= 4.40%, $\chi^2$ = 1.40, $R_{Bragg}$= 8.69%

Unit-cell parameters: a= 5.4763(3), b= 5.4599(3), c= 7.7153(5) Å, V= 230.69(2) Å$^3$.

# Table 2

Fitting Parameters for $La_{0.67}Ca_{0.33}MnO_3$ ($T_{IM}$~180 K)

| T (K) | $\varepsilon_\infty$ | $\Omega_{to}$ ($cm^{-1}$) | $\Omega_{lo}$ ($cm^{-1}$) | $\gamma_{to}$ ($cm^{-1}$) | $\gamma_{lo}$ ($cm^{-1}$) | $S_j$($cm^{-2}$) |
|---|---|---|---|---|---|---|
| | | | $\Omega_{pl}$ ($cm^{-1}$) | $\gamma_0$ ($cm^{-1}$) | $\gamma_{pl}$ ($cm^{-1}$) | |
| | | 163.73 | 166.79 | 52.94 | 137.93 | 0.78 |
| | | 204.24 | 214.17 | 589.03 | 98.01 | 1.05 |
| | | 273.83 | 336.20 | 217.69 | 393.93 | 1.99 |
| 300 | 1.55 | 341.36 | 421.46 | 126.65 | 134.86 | 1.42 |
| | | 580.09 | 610.21 | 79.39 | 94.02 | 0.05 |
| | | 638.23 | 657.75 | 328.25 | 230.15 | 1.88 |
| | | 667.47 | 1133.91 | 725.61 | 2168.36 | 0.71 |
| | | 163.05 | 177..53 | 38.12 | 92.80 | 3.80 |
| | | 211.19 | 224.31 | 1332.73 | 100.12 | 2.09 |
| | | 273.68 | 336.92 | 221.06 | 1413.11 | 5.31 |
| | | 341.13 | 421.15 | 149.00 | 98.92 | 0.20 |
| 190 | 1.57 | 570.28 | 585.72 | 70.92 | 133.63 | 0.58 |
| | | 620.27 | 649.32 | 392.87 | 140.18 | 0.92 |
| | | 668.31 | 1098.97 | 756.88 | 2093.02 | 0.65 |
| | | 163.65 | 165.51 | 33.35 | 57.86 | 0.34 |
| | | 210.58 | 217.58 | 202.62 | 72.04 | 1.45 |
| | | 241.00 | 284.12 | 171.44 | 236.28 | 2.91 |
| 160 | 1.37 | 337.94 | 425.48 | 119.67 | 132.56 | 1.68 |
| | | 578.76 | 611.79 | 77.59 | 79.04 | 0.74 |
| | | 656.71 | 665.18 | 69.09 | 63.42 | 0.38 |
| | | 678.85 | 1077.05 | 900.31 | 2364.04 | 0.67 |
| | | 160.38 | 171.45 | 49.60 | 95.46 | 0.80 |
| | | 202.91 | 212.21 | 401.33 | 114.02 | 2.07 |
| | | 290.79 | 336.17 | 672.58 | 1754.24 | 6.68 |
| 130 | 1.08 | 353.32 | 438.17 | 167.02 | 127.57 | 0.29 |
| | | 578.50 | 590.91 | 70.86 | 100.81 | 0.09 |
| | | 591.15 | 687.02 | 841.44 | 120.50 | 1.98 |
| | | 697.87 | 1224.30 | 197.83 | 2173.34 | 0.48 |
| | | | 939.21 | 601.05 | 610.57 | |
| | | 161.48 | 168.40 | 139.69 | 234.39 | 5.30 |
| | | 190.45 | 210.89 | 283.53 | 127.92 | 6.16 |
| | | 421.49 | 453.51 | 201.34 | 94.30 | 8.87 |
| 77 | | 502.47 | 526.31 | 94.94 | 95.29 | 4.61 |
| | 1.05 | 575.01 | 607.15 | 101.68 | 187.19 | 6.95 |
| | | 627.27 | 720.26 | 1764.57 | 79.20 | 4.43 |
| | | 723.33 | 2955.87 | 109.43 | 3760.98 | 2.99 |
| | | | 1389.09 | 24.96 | 922.12 | |

# Table 3

Fitting Parameters for $La_{0.67}Ca_{0.33}MnO_3$ (Tc ~254 K)

| T (K) | $\varepsilon_\infty$ | $\Omega_{to}$ (cm$^{-1}$) | $\Omega_{lo}$ (cm$^{-1}$)<br>$\Omega_{pl}$ (cm$^{-1}$) | $\gamma_{to}$ (cm$^{-1}$)<br>$\gamma_0$ (cm$^{-1}$) | $\gamma_{lo}$ (cm$^{-1}$)<br>$\gamma_{pl}$ (cm$^{-1}$) | $S_j$(cm$^{-2}$) |
|---|---|---|---|---|---|---|
| 300 | 2.79 | 98.11 | 125.95 | 82.38 | 95.96 | 10.42 |
| | | 173.45 | 203.33 | 47.33 | 329.38 | 3.41 |
| | | 205.03 | 213.55 | 139.05 | 65.70 | 0.04 |
| | | 333.96 | 378.37 | 60.23 | 221.27 | 1.88 |
| | | 389.31 | 398.63 | 39.79 | 61.16 | 0.06 |
| | | 576.10 | 610.21 | 53.40 | 126.80 | 0.71 |
| | | 638.24 | 657.75 | 455.82 | 618.43 | 0.13 |
| | | 2838.10 | 3418.73 | 2134.17 | 2936.99 | 0.98 |
| | | 7985.36 | 8815.68 | 9383.89 | 9804.77 | 1.16 |
| | | | 327.51 | 173.78 | 164.73 | |
| 250 | 2.67 | 98.11 | 125.95 | 70.59 | 95.19 | 8.59 |
| | | 173.45 | 203.33 | 49.97 | 360.151 | 2.81 |
| | | 205.03 | 213.55 | 92.24 | 49.50 | 0.03 |
| | | 333.96 | 378.37 | 61.91 | 238.25 | 1.57 |
| | | 388.31 | 398.63 | 44.28 | 64.23 | 0.05 |
| | | 576.10 | 610.21 | 48.10 | 124.85 | 0.58 |
| | | 638.23 | 657.75 | 483.25 | 664.62 | 0.11 |
| | | 2675.12 | 3274.08 | 1941.45 | 2928.17 | 1.54 |
| | | 7931.93 | 8715.68 | 8415.28 | 9928.75 | 0.39 |
| | | | 340.07 | 154.84 | 127.47 | |
| 200 | 2.19 | 95.65 | 127.33 | 115.12 | 1023.95 | 10.30 |
| | | 176.15 | 203.33 | 96.50 | 545.26 | 3.27 |
| | | 205.03 | 230.76 | 559.92 | 328.06 | 0.10 |
| | | 371.70 | 400.31 | 120.19 | 328.17 | 1.29 |
| | | 405.77 | 439.10 | 360.07 | 431.80 | 0.12 |
| | | 626.21 | 635.42 | 70.20 | 62.42 | 0.27 |
| | | 640.24 | 664.41 | 161.40 | 327.67 | 0.09 |
| | | 2671.78 | 3383.33 | 2264.34 | 3483.63 | 1.66 |
| | | 8053.23 | 8953.10 | 6525.18 | 7595..36 | 0.50 |
| | | | 1006.53 | 104.56 | 106.76 | |
| | | 72.54 | 86.28 | 66.46 | 1702.92 | 5.50 |
| | | 173.11 | 209.35 | 134.47 | 291.75 | 6.24 |
| | | 210.49 | 256.75 | 178.52 | 301.89 | 0.10 |
| 160 | 1.81 | 367.47 | 403.41 | 190.89 | 295.49 | 1.05 |
| | | 406.82 | 441.00 | 121.96 | 271.68 | 0.05 |
| | | 637.36 | 642.86 | 173.87 | 657.82 | 0.23 |
| | | 643.95 | 672.07 | 309.83 | 228.98 | 0.04 |
| | | 2843.96 | 3467.88 | 2076.80 | 2850.51 | 1.14 |
| | | 7988.86 | 9027.79 | 6428.89 | 8205.96 | 0.46 |
| | | | 2015.61 | 222.93 | 227.13 | |
| | | 72.54 | 86.28 | 284..45 | 2123.51 | 6.77 |
| | | 173.11 | 209.35 | 138.43 | 1149.91 | 6.56 |
| | | 210.49 | 256.75 | 447.52 | 873.31 | 0.11 |
| 110 | 1.54 | 367.47 | 413.41 | 873.39 | 431.61 | 1.67 |
| | | 421.02 | 518.94 | 1126.61 | 242.96 | 0.18 |
| | | 588.87 | 590.73 | 182.64 | 92.53 | 0.02 |
| | | 601.30 | 644.79 | 95.51 | 1107.73 | 0.12 |
| | | 3014.04 | 3611.06 | 1904.02 | 2516.03 | 0.72 |
| | | 9082.81 | 9519.18 | 6878.62 | 7473.40 | 0.14 |
| | | | 3261.02 | 380.01 | 421.15 | |
| | | 72.54 | 86.28 | 59.85 | 2188.07 | 5.44 |
| | | 173.45 | 209.34 | 154.73 | 591.04 | 6.16 |
| | | 210.49 | 256.75 | 214.64 | 183.10 | 0.09 |
| 77 | | 367.46 | 403.92 | 205.41 | 246.84 | 1.05 |
| | 1.20 | 406.82 | 445.00 | 97.32 | 236.86 | 0.04 |
| | | 588.07 | 590.70 | 245.35 | 79.42 | 0.03 |
| | | 601.10 | 604.42 | 72.73 | 726.81 | 0.02 |
| | | 2813.27 | 4005.99 | 2746.03 | 4039.63 | 1.89\ |
| | | 8050.30 | 9799.15 | 7904.12 | 9996.06 | 0.50 |
| | | | 3829.10 | 315.57 | 312.96 | |

## TABLE 4

Parameters of the small polaron theory for $La_{0.67}Ca_{0.33}MnO_3$ (T~180 K)

| T (K) | $\eta_1$ | $\varpi_{ph1}$ ($cm^{-1}$) | $\eta_2$ | $\varpi_{ph2}$ ($cm^{-1}$) | $\eta_3$ | $\varpi_{ph3}$ ($cm^{-1}$) | $\rho$ (ohm-cm) |
|---|---|---|---|---|---|---|---|
| **300** | 11.46 | 172.0 | 5.90 | 645. | 6.90 | 1388 | 0.1532 |
| **190** | 11.45 | 168. | 6.48 | 645.2 | 8.11 | 1228.5 | 0.2146 |
| **160** | 8.05 | 216.90 | 6.49 | 588.20 | 8.28 | 1097.51 | 0.1906 |
| **130** | 7.35 | 161.0 | 5.1 7 | 399.84 | 6.27 | 668.00 | 0.0962 |
| **77** | 7.38 | 166.50 | 6.49 | 369.5 | 8.0 | 670.00 | 0.0882 |

Note that for 130 K it is necessary to add: $\eta_4$=8.57, $\varpi_{ph4}$=1133.42 while for 77 K $\varpi_{ph4}$=1300, $\eta_4$=12.29.

## TABLE 5

Parameters of the small polaron theory for $La_{0.67}Ca_{0.33}MnO_3$ (T~ 254 K).

| T (K) | $\eta_1$ | $\varpi_{ph1}$ ($cm^{-1}$) | $\eta_2$ | $\varpi_{ph2}$ ($cm^{-1}$) | $\eta_3$ | $\varpi_{ph3}$ ($cm^{-1}$) | $\eta_4$ |
|---|---|---|---|---|---|---|---|
| **300** | 24.4 | 141.70 | 6.5<br>6.7 | 620.5<br>540.3 | 7.59 | 1190 | 8.3 |
| **250** | 23.1 | 145.7 | 6.1<br>6.55 | 645.<br>583. | 8.15 | 1125 | 10.25 |
| **200** | 9.6 | 135.7 | 6.16<br>6.46 | 601.5<br>545. | 8.23 | 1190 | 9.30 |
| **160** | 6.85 | 135.7 | 5.28<br>5.46 | 640.5<br>600. | 8.20 | 1220.1 | |
| **110** | 6.7 | 135.7 | 5.6<br>5.3 | 618<br>611 | 8.55 | 1096 | 9.30 |
| **77** | 2.52 | 270.70 | 4.86 | 572. | 6.62 | 928 | 9.44 |

# TABLE 6

Structural analysis of nearest neighbors oxygen coordination in $La_{0.67}Ca_{0.33}MnO_3$ ($T_{IM}$ ~254K)

| T(K) | N | R(Å) | $\Delta\sigma^2$ |
|---|---|---|---|
| **10** | | | |
| 3-shell | 3.6 ± 0.1 | 1.92±0.02 | -0.0006 |
| | 0.6±0.1 | 2.06±0.02 | -0.0039 |
| | 1.0±0.3 | 2.07±0.03 | -0.0001 |
| 2-shell | 3.4 ± 0.06 | 1.92±0.02 | -0.0014 |
| | 1.3±0.03 | 2.05±0.02 | -0.0035 |
| | | | |
| **90** | | | |
| 3-shell | 3.3 ± 0.1 | 1.92±0.02 | -0.0006 |
| | 0.8±0.1 | 2.04±0.02 | -0.0019 |
| | 0.8±0.3 | 2.04±0.03 | -0.0019 |
| 2-shell | 3.3 ± 0.07 | 1.92±0.02 | -0.0005 |
| | 1.6±0.03 | 2.04±0.02 | -0.0019 |
| | | | |
| **160** | | | |
| 3-shell | 3.4± 0.1 | 1.918±0.02 | -0.0005 |
| | 0.3±0.1 | 1.994±0.02 | -0.002 |
| | 0.9±0.3 | 2.124±0.03 | -0.0001 |
| | | | |
| **250** | | | |
| 3-shell | 3.4 ± 0.1 | 1.92±0.02 | 0.0010 |
| | 0.4±0.1 | 2.05±0.02 | -0.0036 |
| | 0.4±0.1 | 2.15±0.03 | -0.0022 |
| | | | |
| **300** | | | |
| 3-shell | 3.6 ± 0.1 | 1.92±0.02 | 0.0017 |
| | 0.1±0.1 | 2.04±0.02 | 0.0067 |
| | 0.6±0.3 | 2.15±0.03 | 0.0022 |

N, coordination number; $R_i$ interatomic distances and $\Delta\sigma_i$ are Debye-Waller factors relative to the FEFF6 simulation.

## TABLE 7

Structural analysis of nearest neighbors oxygen coordination in $La_{0.67}Ca_{0.33}MnO_3$ intermediate case ($T_{IM1}$~180 K and $T_{IM2}$~254 K).

| T(K) | N | R(Å) | $\Delta\sigma^2$ |
|---|---|---|---|
| **10** | | | |
| 2-shell | 3.4 ± 0.06 | 1.92±0.02 | -0.0007 |
| | 1.4±0.03 | 2.03±0.02 | -0.003 |
| | | | |
| **160** | | | |
| 3-shell | 3.4± 0.1 | 1.92±0.02 | -0.0002 |
| | 0.5±0.1 | 1.97±0.02 | -0.006 |
| | 0.9±0.3 | 2.14±0.03 | -0.005 |

N coordination number; $R_i$ interatomic distances and $\Delta\sigma_i$ are Debye-Waller factors relative to the FEFF6 simulation.

## TABLE 8

Structural analysis of nearest neighbors oxygen coordination in $La_{0.67}Ca_{0.33}MnO_3$ ($T_{IM}$~180 K)

| T(K) | N | R(Å) | $\Delta\sigma^2$ |
|---|---|---|---|
| 10 | | | |
| 3-shell | 3.8 ± 0.1 | 1.92±0.02 | 0.0005 |
| | 0.4±0.1 | 2.00±0.02 | -0.0035 |
| | 1.1±0.4 | 2.10±0.03 | -0.000 |
| 2-shell | 3.7 ± 0.06 | 1.92±0.02 | -0.0004 |
| | 1.7±0.03 | 2.03±0.02 | -0.0019 |
| | | | |
| 90 | | | |
| 3-shell | 3.7 ± 0.1 | 1.92±0.02 | 0.0003 |
| | 0.6±0.1 | 2.00±0.02 | -0.0040 |
| | 0.8±0.3 | 2.13±0.03 | -0.0028 |
| | | | |
| 160 | | | |
| 3-shell | 3.4± 0.1 | 1.92±0.02 | -0.001 |
| | 0.5±0.1 | 1.97±0.02 | -0.005 |
| | 0.8±0.3 | 2.12±0.03 | -0.004 |
| | | | |
| 250 | | | |
| 3-shell | 3.8 ± 0.1 | 1.92±0.02 | 0.0014 |
| | 0.3±0.1 | 2.03±0.02 | -0.006 |
| | 0.6±0.1 | 2.17±0.03 | -0.0020 |
| | | | |
| 300 | | | |
| 3-shell | 3.8 ± 0.1 | 1.92±0.02 | 0.0018 |
| | 0.4±0.1 | 2.03±0.02 | -0.003 |
| | 0.3±0.3 | 2.15±0.03 | -0.003 |

N, coordination number; $R_i$ interatomic distances and $\Delta\sigma_i$ are Debye-Waller factors relative to the FEFF6 simulation.

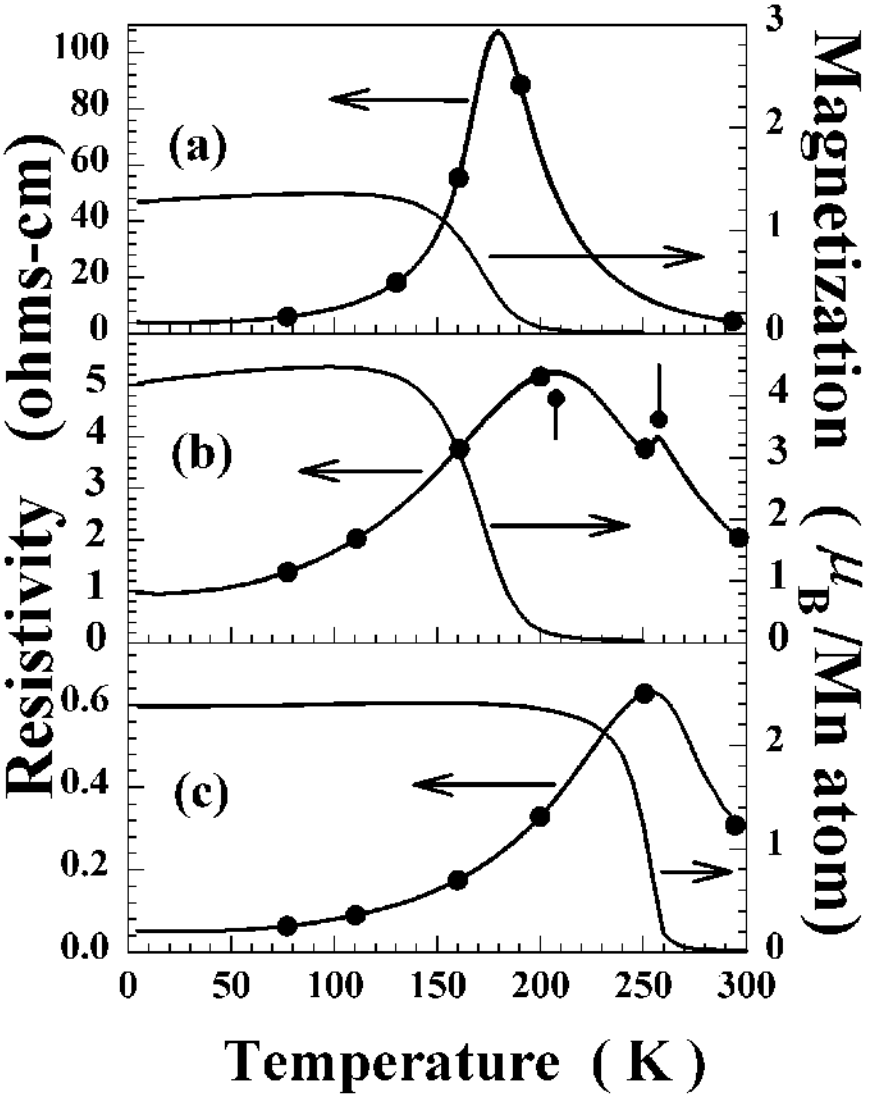

Figure 1
Massa et al

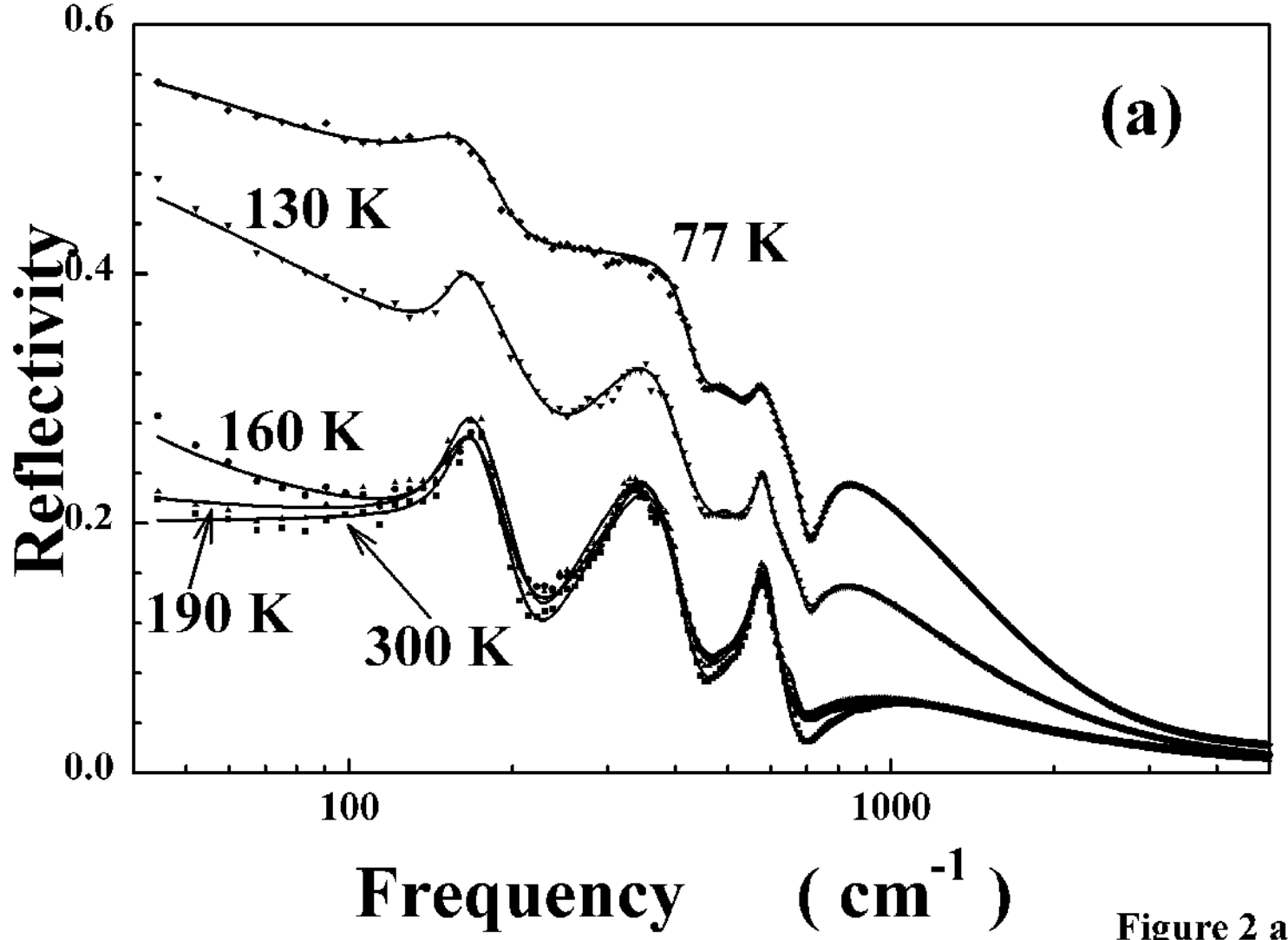

Figure 2 a
Massa et al

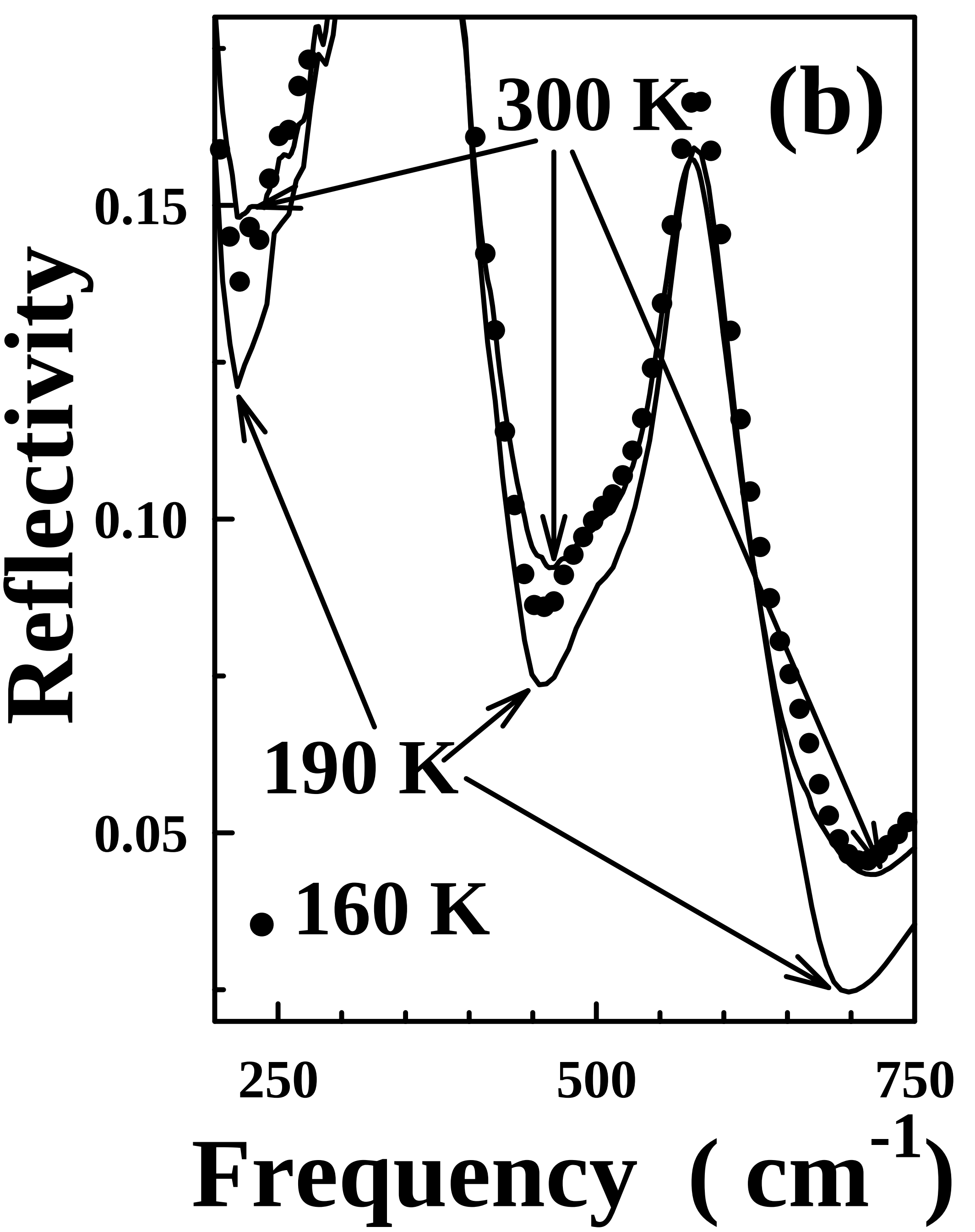

Figure 2 b
Massa et al

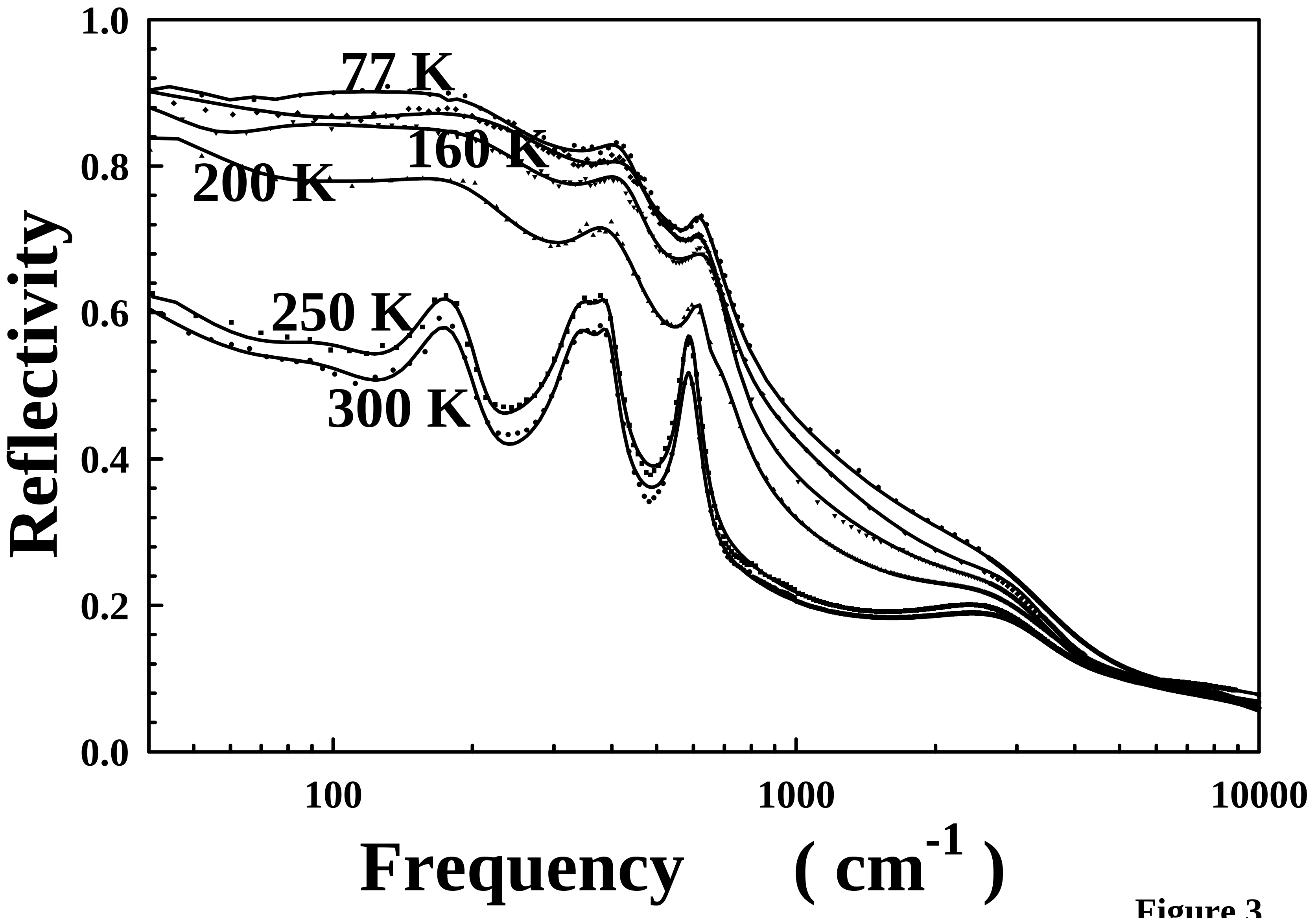

**Figure 3**
**Massa et al**

**Figure 4**
**Massa et al**

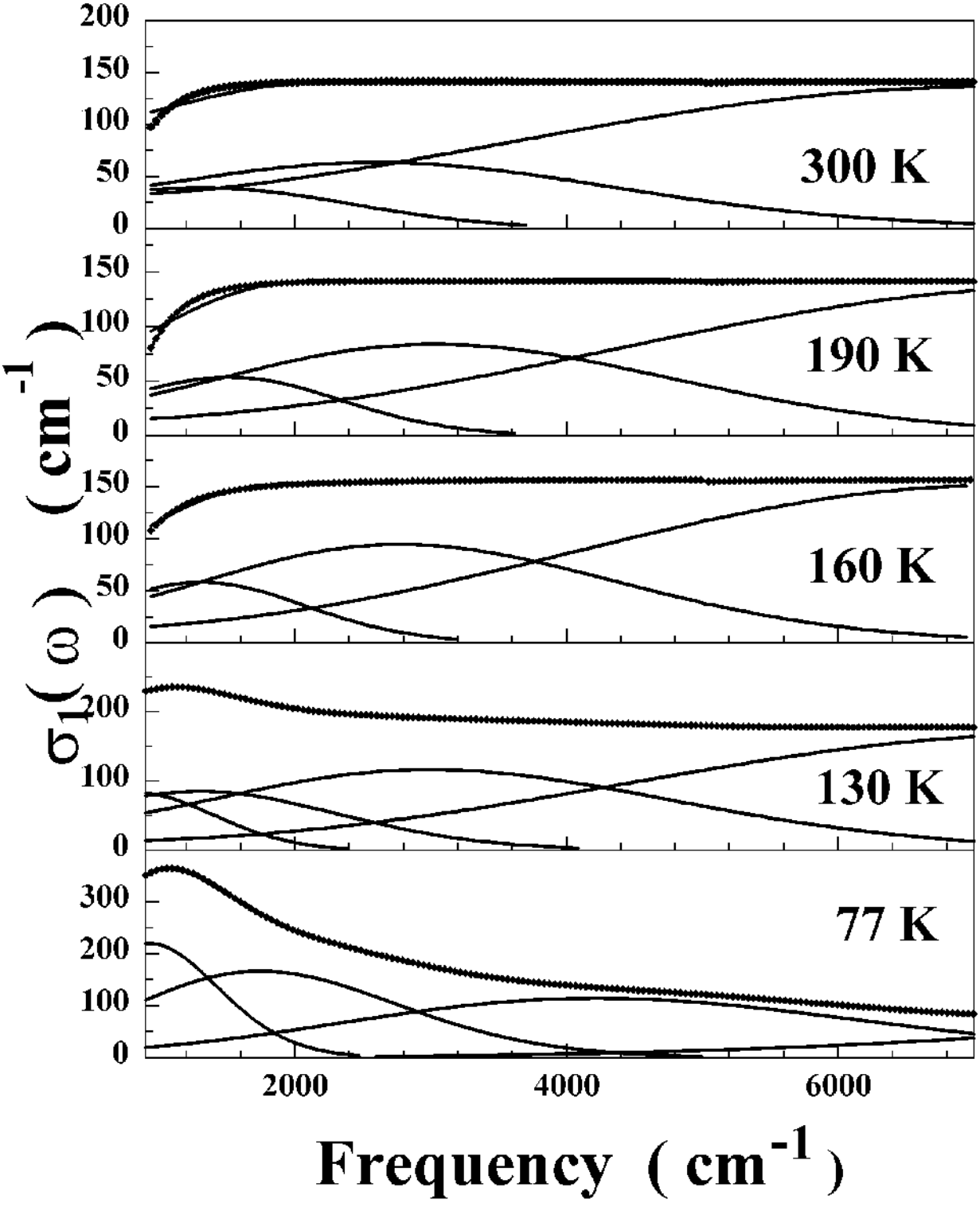

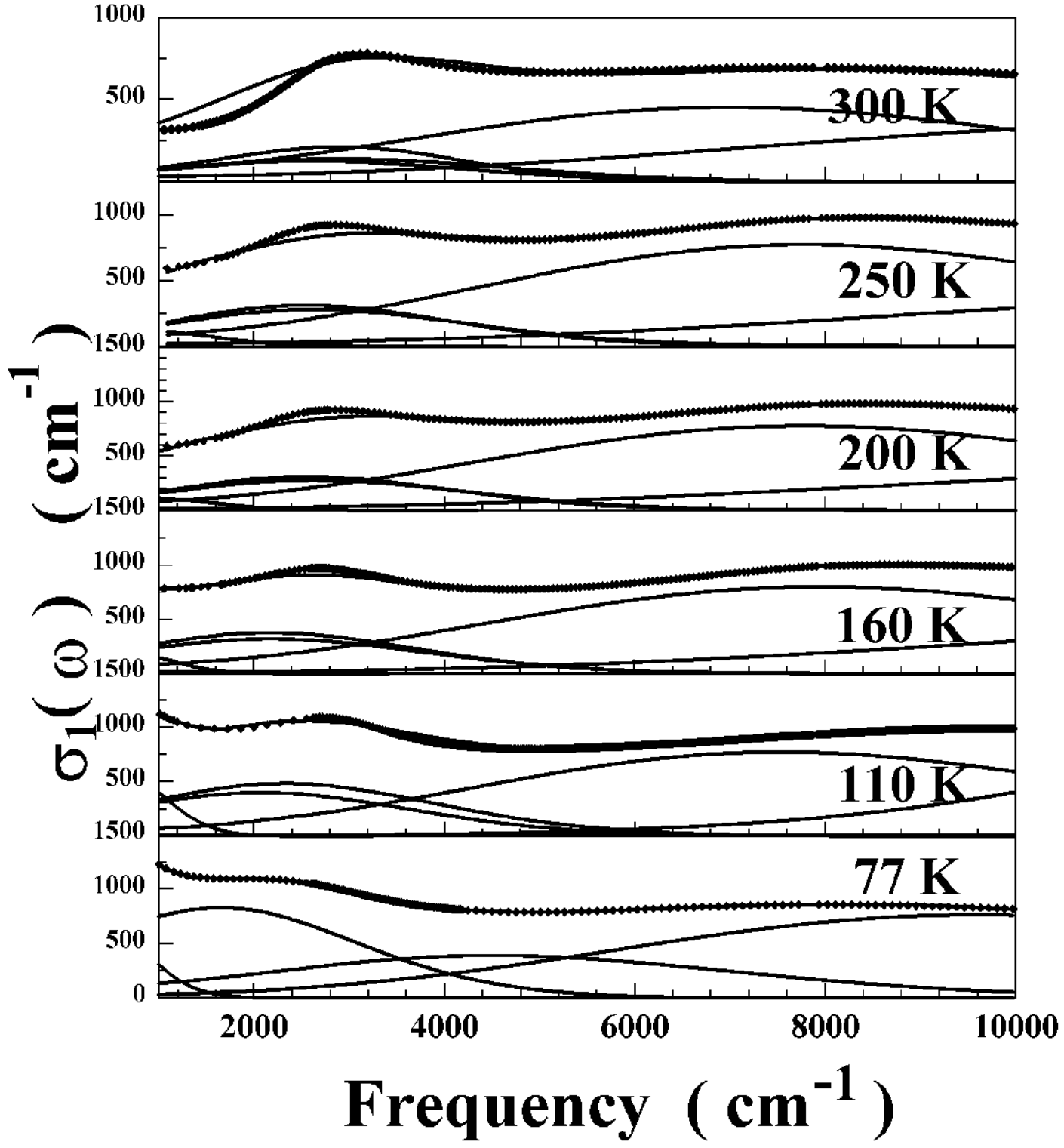

**Figure 5**
**Massa et al**

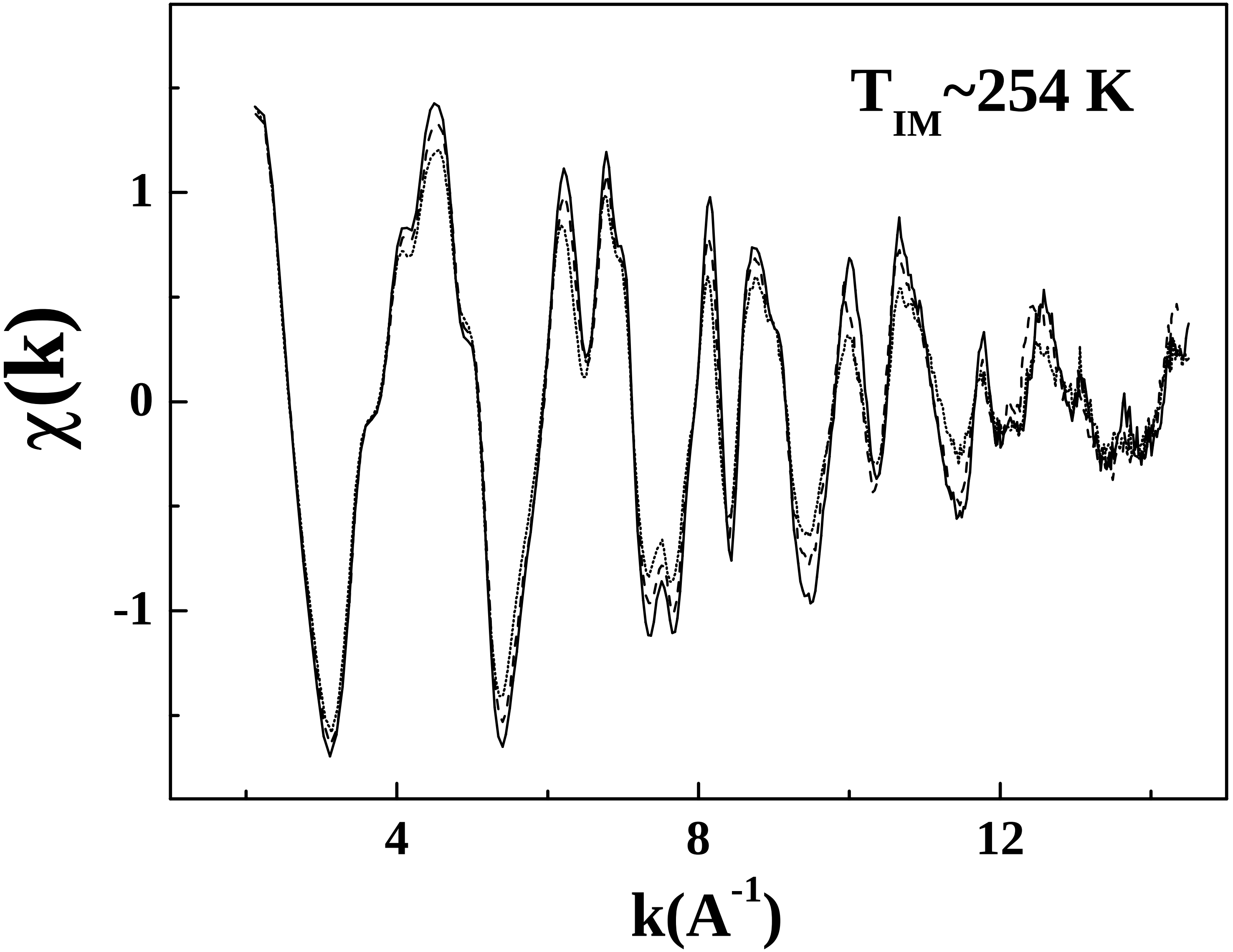

**Figure 6**
**Massa et al**

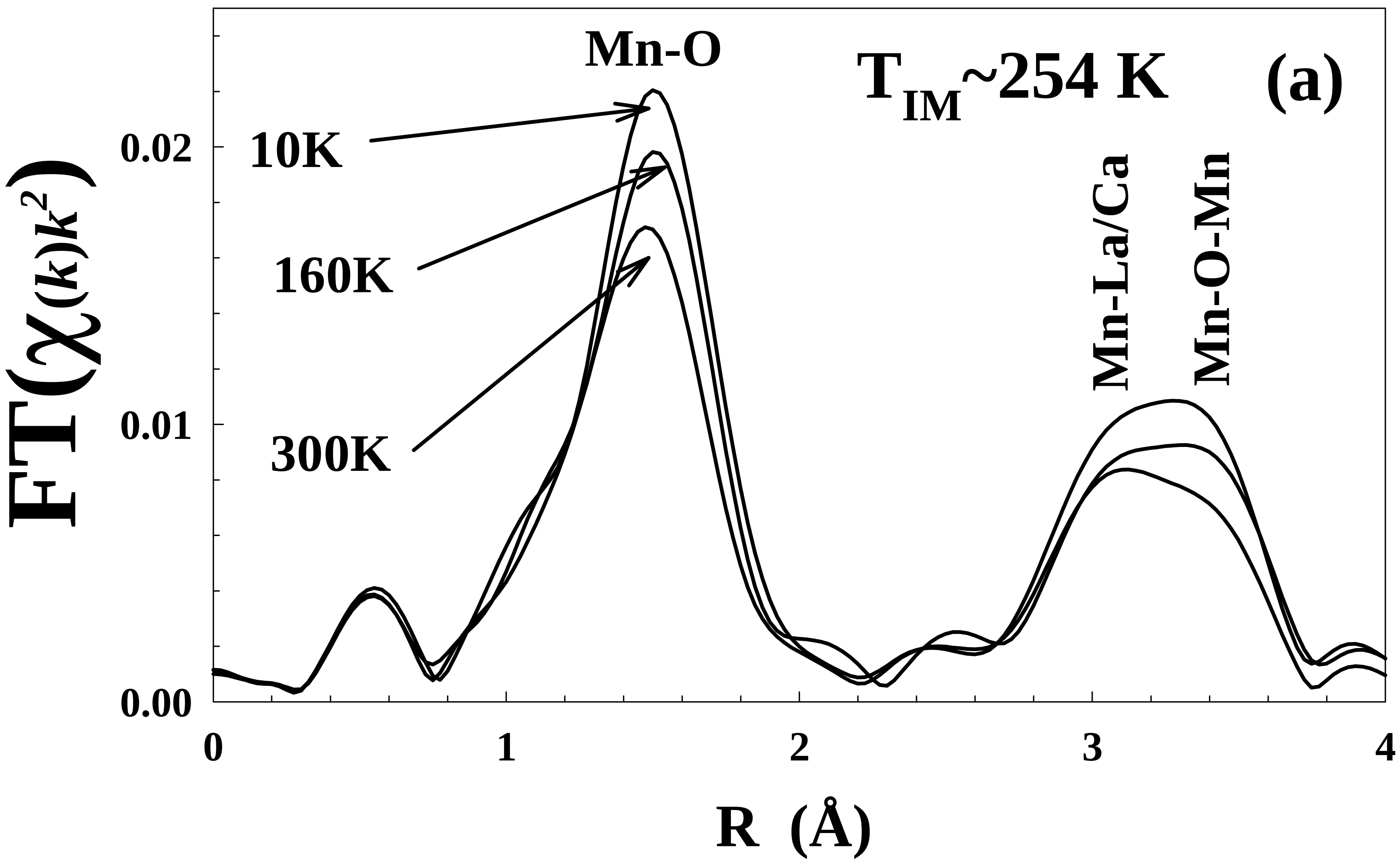

**Figure 7 a**
**Massa et al**

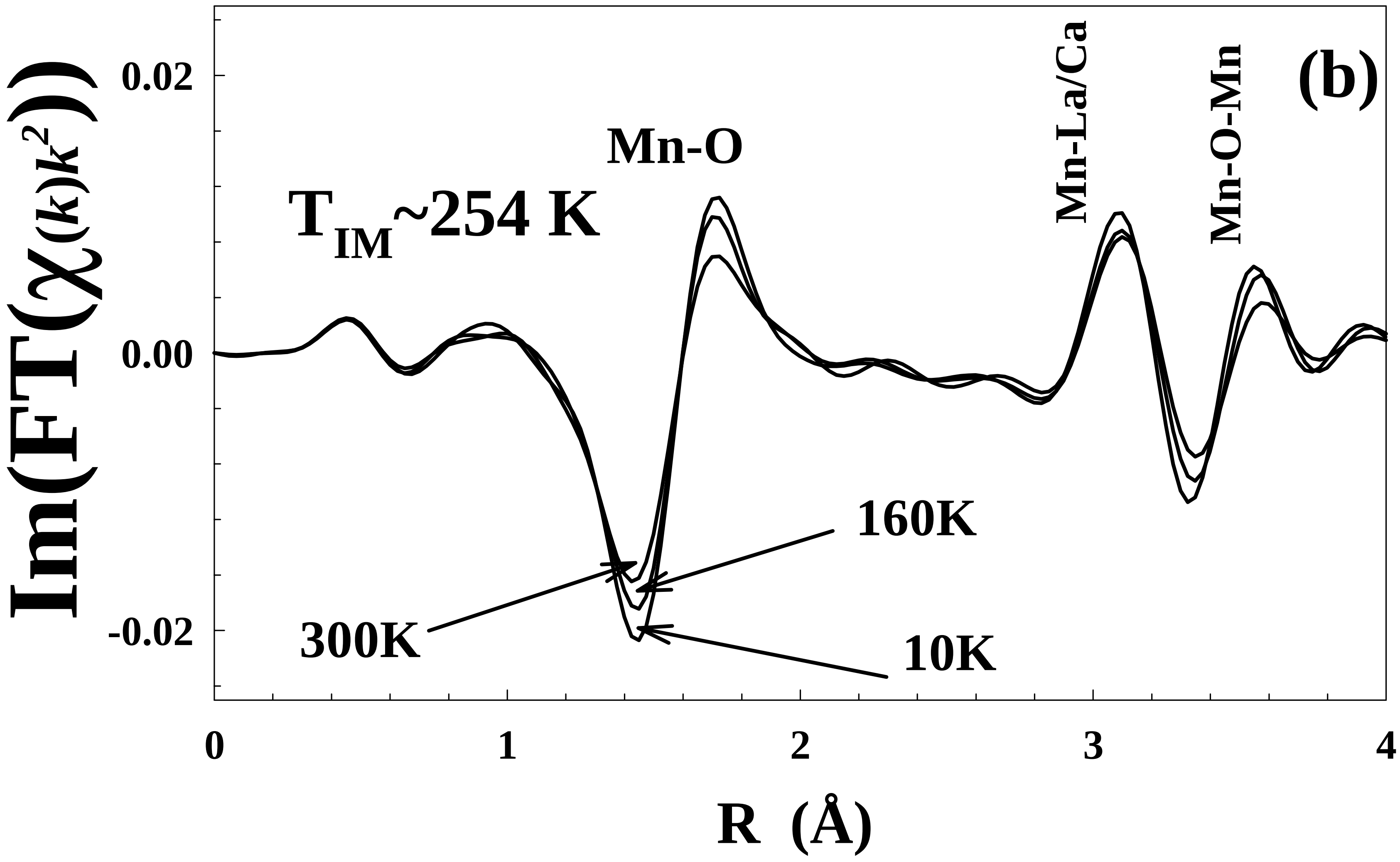

Figure 7 b
Massa et al

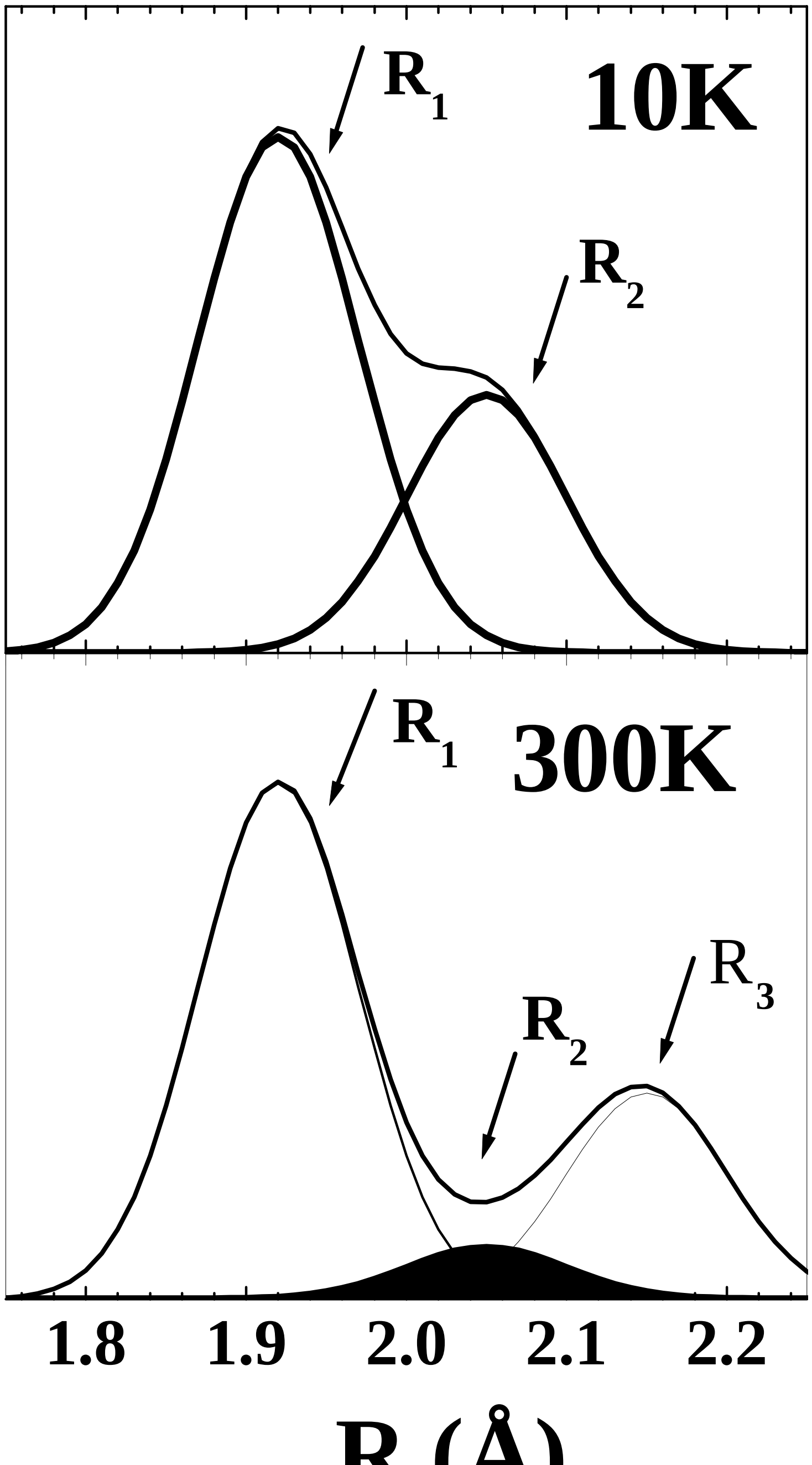

**Figure 8**
**Massa et al**

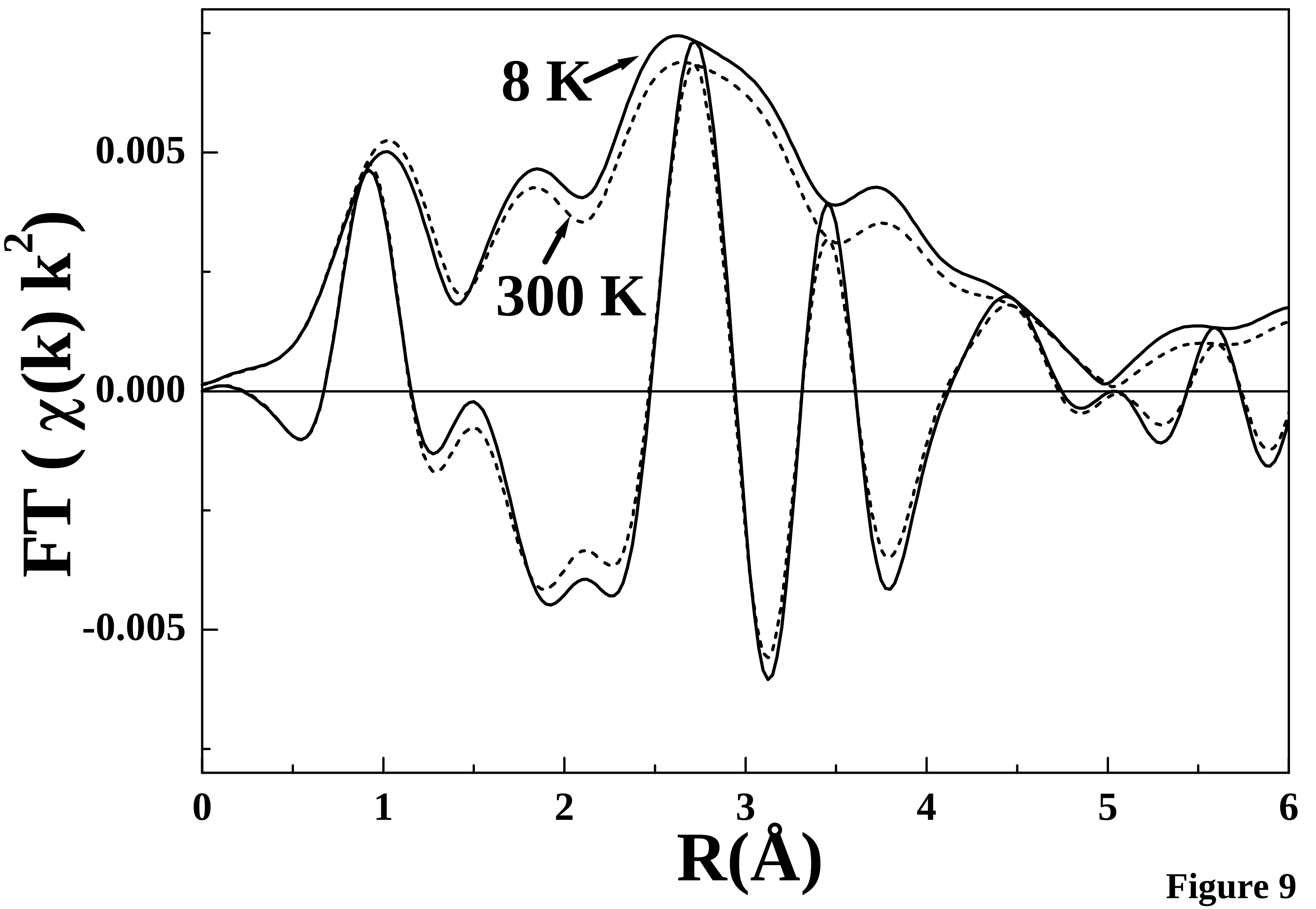

**Figure 9**
**Massa et al**